# 24 Saturn's Exploration Beyond Cassini-Huygens

Tristan Guillot
Observatoire de la Côte d'Azur, CNRS

Sushil Atreya
University of Michigan, Ann Arbor

Sébastien Charnoz
Equipe AIM , CEA/SAp, Université Paris Diderot

Michele K. Dougherty
Imperial College London, SW7 2AZ, UK

Peter Read
University of Oxford

## Abstract

*For its beautiful rings, active atmosphere and mysterious magnetic field, Saturn is a fascinating planet. It also holds some of the keys to understanding the formation of our Solar System and the evolution of giant planets in general. While the exploration by the Cassini-Huygens mission has led to great advances in our understanding of the planet and its moons, it has left us with puzzling questions: What is the bulk composition of the planet? Does it have a helium core? Is it enriched in noble gases like Jupiter? What powers and controls its gigantic storms? We have learned that we can measure an outer magnetic field that is filtered from its non-axisymmetric components, but what is Saturn's <u>inner</u> magnetic field? What are the rings made of and when were they formed?*

*These questions are crucial in several ways: a detailed comparison of the compositions of Jupiter and Saturn is necessary to understand processes at work during the formation of these two planets and of the Solar System: was the protosolar disk progressively photoevaporated of its hydrogen and helium while forming its planets? Did Jupiter and Saturn form at the same time from cores of similar masses? Saturn is also a unique laboratory for studying the meteorology of a planet in which, in contrast to the Earth, the vapor of any condensing species (in particular water) is <u>heavier</u> than the surrounding air. A precise measurement of its magnetic field is needed to constrain dynamo theories and apply it to other contexts, from our Earth to extrasolar planets. Finally, the theory behind the existence of its rings is still to be confirmed, and has consequences for a variety of subjects from theories of accretion of grains to the study of physical mechanisms at work in protoplanetary systems.*

*All in all, this calls for the continued exploration of the second largest planet in our Solar System, with a variety of means including remote observations and space missions. Measurements of gravity and magnetic fields very close to the planet's cloud tops would be extremely valuable. Very high spatial resolution images of the rings would provide details on their structure and the material that form them. Last but not least, one or several probes sent into the atmosphere of the planet would provide the critical measurements that would allow a detailed comparison with the same measurements at Jupiter.*

## 24.1 Introduction

Saturn was probably first observed with a telescope by Galileo in 1610 but it was not until 1655 that Christiaan Huygens discovered its largest moon, Titan. Four years later, he correctly inferred that the planet has rings. Then, between 1671 and 1684, Jean-Dominique Cassini discovered Saturn's moons Japetus, Rhea, Thethys and Dione, as well as the now so-called Cassini division. Although the planet fascinated many, the following major milestones in the discovery of this world had to await the first space missions, Pioneer 11 in 1977, Voyager 1 in 1980 and Voyager 2 in 1981. Among many results, the missions measured the planet's atmospheric composition, discovered Saturn's magnetic field, measured Saturn's wind patterns. They gave evidence of the amazing thinness of the rings and of their structure. Then in 2004 came the Cassini-Huygens spacecraft which, to list but a few results, further extended our knowledge of Saturn's system of moons, composition of the rings, and unveiled Saturn's meteorology in all its complexity.

Saturn is a truly major planet in the Solar System: with 95 times the mass of the Earth, it is the second largest planet. It contains a large fraction of hydrogen and helium, gases that were most abundant when the Solar System was formed. As such, it is a witness of events that occurred very early during the formation of the system, for which the study of its formation provides us with invaluable information. About 800 million years later, it was probably responsible for a reorganization of the system of outer planets and Kuiper-Edgeworth belt, which led to the so-called "late heavy bombardment" in the inner Solar System. Saturn also has a complex atmosphere, both in terms of chemistry and dynamics and, as such, it is a fantastic laboratory for understanding planetary atmospheres in general. Among the planets in the Solar System, its magnetic field is second only to Jupiter in intensity, and it has an unusual, completely axisymmetric form that is still unexplained. Its rings are a laboratory for understanding disks and can be sought as miniature protoplanetary systems. Finally, in the era of the discovery of planets, and particularly giant planets, around other stars than our Sun, understanding Saturn's thermal evolution is crucial.

In spite of all the efforts and progress made in the past 30 years or so, Saturn remains mysterious. We will review the many questions, some of them unexpected, that the Cassini-Huygens mission has left us with and that call for a continued exploration of this planet. First, we will discuss how the interior of the planet remains uncertain. We will then see how understanding the evolution, and hence the composition, of giant exoplanets is tied to understanding the evolution of Saturn. The next sections will discuss Saturn's atmospheric composition, atmospheric dynamics, and rings respectively. We will then see how a better understanding of the planet would help us in deciphering the mystery of the origin of our Solar System. The means to explore Saturn further during the next several decades will be discussed.

## 24.2  Saturn's interior

Saturn is known to be mostly made of hydrogen and helium but to contain other elements ("heavy elements") in a proportion that is significantly larger than is the case in the Sun. It is also known to contain a central dense core of 10 to 25 Earth masses that was probably the seed of the formation of the planet, before it accreted its hydrogen-helium envelope from the protosolar disk (e.g. Guillot 2005). *Qualitatively*, these conclusions have stood almost unchanged for more than twenty five years (e.g. Stevenson 1982), although *quantitative* improvements are due to better equations of state, and improvements in computing power. These inferences have rested on the calculation of interior models fitted to several key measurements: the planet's mass, radius, atmospheric temperature and pressure, atmospheric helium abundance, interior rotation rate and gravity field (gravitational moments $J_2$, $J_4$ and $J_6$). An important astrophysical constrain is the primordial (protosolar) helium to hydrogen ratio, with the hypothesis that missing helium in the planet's atmosphere is hidden deeper into the planet. A final, crucial ingredient is an equation of state, most importantly for hydrogen, the most abundant element, but ideally for all species to be considered.

While Cassini-Huygens has provided a much better measurement of the gravity field of the planet by an order of magnitude, it has also demonstrated that the inner rotation rate is much less constrained than it was thought to be. Furthermore, uncertainties have remained as to whether Saturn's atmosphere is very poor, or moderately poor in helium, with a mass fraction in helium still ranging between 7% and 17% (28% being the protosolar value). In parallel, while progresses have been made with high-pressure experiments in the Mbar regime, both in the lab and numerically, hydrogen has proven to be a surprisingly difficult substance to comprehend. As a result, several theories exist to explain several experiments, but they generally do not agree with each other! Naturally, the case of the hydrogen-helium mixture has remained even more difficult and no reliable prediction exist as to the temperature-pressure regime in which hydrogen and helium may separate out into two mixtures of differing compositions in the giant planets. Yet, this is crucial for explaining the missing atmospheric helium, as such a phase separation would yield the rapid formation of helium-rich droplets that would fall towards the deeper regions (Stevenson & Salpeter 1977) with consequences for the planet's interior structure.

*Figure 1* shows two possible structures for Saturn's interior: in the traditional view, the helium phase separation occurs close to the molecular-metallic transition. The planet is then thought to consist of a helium poor molecular hydrogen envelope, a helium-rich metallic hydrogen envelope and a central dense core of unknown composition (e.g. Saumon & Guillot 2004). Alternatively, helium sedimentation may proceed all the way to the core in which case a helium shell may be present on top of the central core in which case most of the envelope should be helium-poor (Fortney & Hubbard 2003). It is also important to note that differential rotation in the interior has been traditionally neglected, but may play an important role (Hubbard 1999; see Chapter by Hubbard et al.).

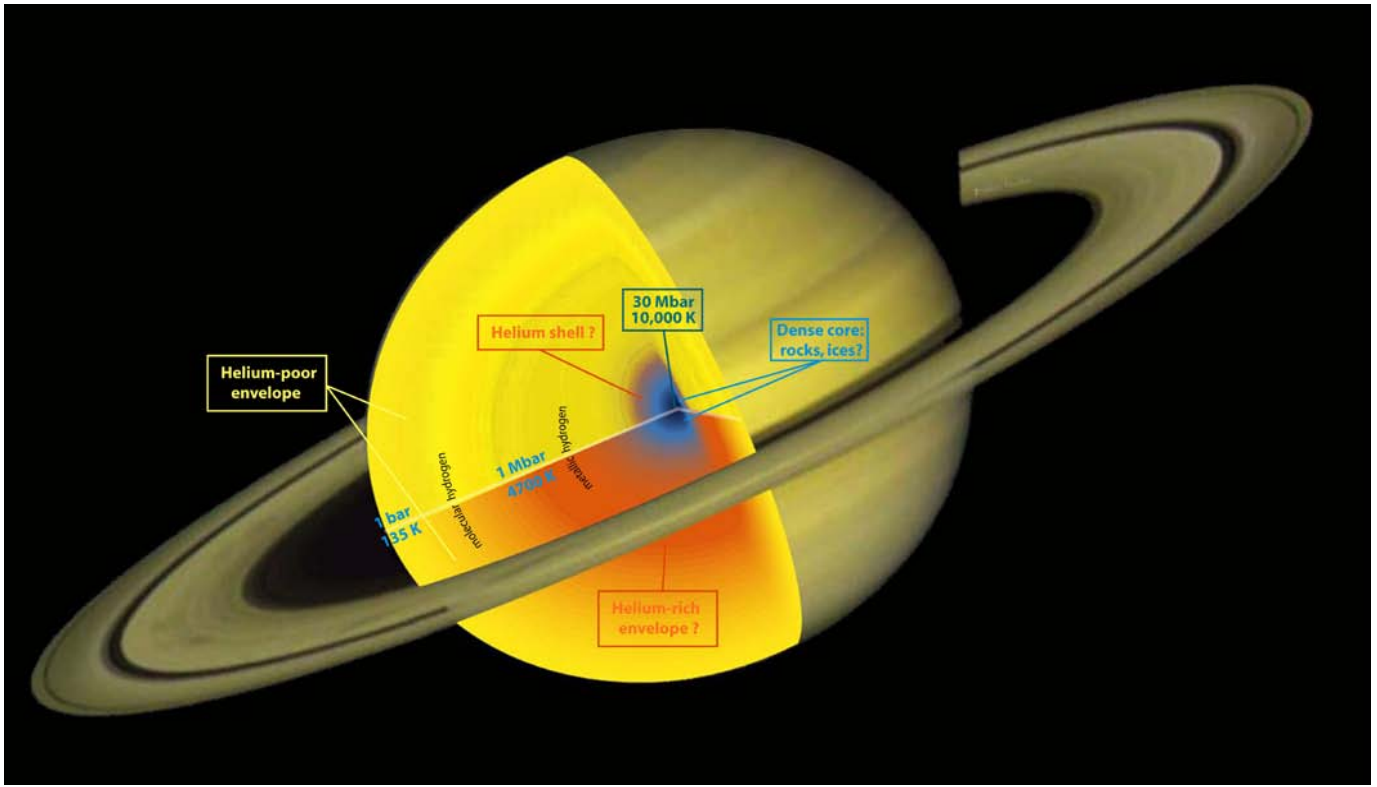

*Figure 1: Sketch of the interior of Saturn showing two possible classes of models: (1) Bottom: The envelope is split in a helium-poor region in which hydrogen is in molecular form and a helium-rich region with metallic hydrogen. (2) Top: Helium sedimentation is supposed to be complete down to the core so that a helium shell is formed at the bottom of a helium-poor envelope. In both cases, a central core of made of dense material (most probably a mixture of iron, rocks, ices in unknown proportion and state) is supposed to be present.*

Altogether, progressing in our knowledge of Saturn's interior will require:
- To determine accurately the abundance of helium in Saturn's atmosphere.
- To obtain a reliable hydrogen-helium equation of state and phase diagram in the relevant pressure-temperature range (from 0.1 to 30 Mbar, and 1000 to 20000K) and for mixtures of hydrogen, helium and heavy elements (most importantly water).
- To measure the deep water abundance (water is extremely important both because it forms almost half of the mass of a solar composition mixture of heavy elements, and because it directly impacts the planet's meteorology).
- To measure accurately Saturn's interior rotation rate.
- To obtain accurate measurements of Saturn's gravity field including high order gravitational moments.
- With numerical experiments, to model Saturn's interior rotation to determine how atmospheric zonal winds and interior rotation are linked, with the aim of being able to invert gravitational field measurements for accurate constraints on the density profile.

- To model the extent to which elements can be mixed upward or inversely settle towards central regions as a function of their behaviors at high pressures.

Progresses in our knowledge of Saturn's interior should parallel that on Jupiter, the comparison between the two planets being extremely instructive in terms of constraints on their interior structures and formation (see §0).

## 24.3 Structure and evolution of low-density giant planets

The discovery that extrasolar giant planets abound in the Galaxy and the possibility to directly characterize them especially when they transit in front of their star is a great opportunity to know more about these planets and the formation of planetary systems (e.g. Charbonneau et al. 2006). Different Saturn-mass extrasolar planets have been found: HD149026b (Sato et al. 2006) has a mass of 1.2 $M_{Saturn}$ (114 $M_\oplus$) for a radius of 0.9 $R_{Saturn}$, implying that it contains about 70 $M_\oplus$ of heavy elements (e.g. Ikoma et al. 2006)! On the other side of the spectrum, HAT-P-1b has a mass of 1.8 $M_{Saturn}$ (170 $M_\oplus$) for a radius of 1.4 $R_{Saturn}$ (Bakos et al. 2007), implying that it contains little heavy elements (Guillot 2008). However, most of the knowledge that can be gained on these objects that lie many tens of light years away rests on what we have learned on the structure and evolution of giant planets in our Solar System. Details such as the influence of atmospheric winds on the structure and cooling, the interior rotation of the planets, the presence of a dynamo, the extent of a phase separation of elements in the planet…etc., all must rest on models tuned to reproduce the structure and evolution of giant planets in our system for which minute details are known.

It is important in this respect to stress that the known 4.56 Ga age of giant planets in our Solar System is unfortunately not yet precisely accounted for. While Jupiter models traditionally reach that value to within 10% (which in itself should be improved), the situation is dire for Saturn for which traditional models fall short of this value by 2 to 2.5 Ga (e.g. Hubbard et al. 1999). This is most probably due to the presence of the helium-hydrogen phase separation at high pressures and to the subsequent release of gravitational energy as the helium-rich droplets fall towards the planetary interior. A model age of ~4.5 Ga can be obtained by accounting for this extra energy source and properly tuning the H/He phase diagram. However there is yet no model that properly accounts for the ages of both Jupiter and Saturn at the same time. The application to extrasolar planets is to be made with caution (see Fortney & Hubbard 2004).

It is therefore most important that the evolution of the two giant planets closest to us is better understood, in order to apply this knowledge to the ensemble of extrasolar giant planets found so far. This requires using better equations of state including phase diagrams of the major species and their mixtures, updated atmospheric models, and an improved understanding of differential rotation and mixing in the planet's interior.

## 24.4 Saturn's atmospheric composition

Saturn's atmosphere has been investigated by remote sensing from the ground, earth orbit (HST, ISO) and spacecraft (Pioneer 11, Voyager 1 and Voyager 2, Cassini). As a result, much is known about the composition of Saturn's stratosphere. Column abundances of various hydrocarbons, including $CH_4$, $CH_3$, $C_2H_2$, $C_2H_4$, $C_2H_6$, $C_3H_8$, $C_4H_2$, $C_6H_6$, oxygen-bearing molecules, including $H_2O$, $CO$, $CO_2$, disequilibrium species, including $PH_3$, $GeH_4$, $AsH_3$ (and CO), and certain other minor constituents such as $NH_3$ and HCl, have been determined (Atreya et al., 1999). However, their vertical density distributions have not been measured, with the exception of $CH_4$ and $C_2H_2$ whose profiles were derived over a couple of scales heights in the vicinity of Saturn's homopause using the solar occultation technique in the ultraviolet from Voyager. No composition or temperature information is available for the middle atmosphere, the "ignorosphere", extending roughly from 1 mb to 1 nanobar. Furthermore, composition in the troposphere has not been determined except for methane. Being the principal reservoir of carbon at Saturn, methane provides the carbon elemental abundance. The stratospheric hydrocarbons are photochemical products of methane, and thus inappropriate for deriving the C/H.

Both Voyager and Cassini spacecraft were able measure the abundance of methane in the well-mixed troposphere. The high precision data from the Cassini Composite Infrared Spectrometer (CIRS) have yielded a $CH_4$ mole fraction of 4.5 ± 0.9 x $10^{-3}$, or $CH_4/H_2$ = 5.1 ± 1.0 x $10^{-3}$ (Flasar et al., 2005). This implies the carbon elemental ratio, C/H = 9.3±1.8 x solar, using the Grevesse et al. (2005) solar elemental abundances. CIRS also determined P/H=10 x solar from $PH_3$ in the upper troposphere (Fletcher et al., 2007). Although similar to C/H, it may not represent the true deep, well-mixed atmosphere elemental abundance of phosphorus, as $PH_3$ is a disequilibrium species which is in thermochemical equilibrium at several thousand bars where the temperatures are several thousand degrees Kelvin. Finally, ground-based VLA observations in the microwave have yielded rather uncertain, model-dependent results for N/H=2-4 x solar from $NH_3$ and S/H=12 x solar from $H_2S$ (Briggs and Sackett, 1989). In summary, the only heavy element with reliable data on its abundance in Saturn's atmosphere is carbon.

Whereas the Galileo entry probe measured the bulk atmospheric composition of Jupiter except for water, the bulk composition of Saturn's atmosphere, hence the abundance of the heavy elements, remains mysterious for the most part. The planet's well-mixed atmosphere is representative of the bulk composition, but it lies below the clouds, especially for condensible species (NH3, H2S, H2O in Saturn and Jupiter). Remote sensing, such as that from the Cassini orbiter, is not a suitable technique for sampling this part of the atmosphere. On the other hand, elemental abundances, especially those of the heavy elements are required to constrain the models of the formation of Saturn and the origin of its atmosphere. The most critical heavy elements are O, C, N, S and the noble gases Ne, Ar, Kr, Xe. In addition, isotope abundances of the noble gases, D/H, $^3He/^4He$, $^{14}N/^{15}N$ are also important for gaining an insight into the origin and evolution of the atmosphere. Finally, a precise determination of the He/H ratio in the atmosphere provides a window into interior processes such as the conversion of helium into liquid metallic form in the 3-5 megabar region (as predicted from the equation of state and laboratory experiments), release of gravitational potential energy, etc. The revised Voyager analysis yielded $He/H_2$ = 0.11-17 (Conrath and Gautier, 2000), which has too large an uncertainty to be of value in discriminating between various models of the interior of Saturn. Direct in situ measurements of the helium abundance are needed, as was done at Jupiter by the Galileo probe. Moreover, determination of the elemental and isotope abundances requires accessing and measuring the well-mixed part of the atmosphere for a large number of key constituents. The pressure levels at which the atmosphere can be considered as well-mixed depend on the constituent, as illustrated in *Figure 2*.

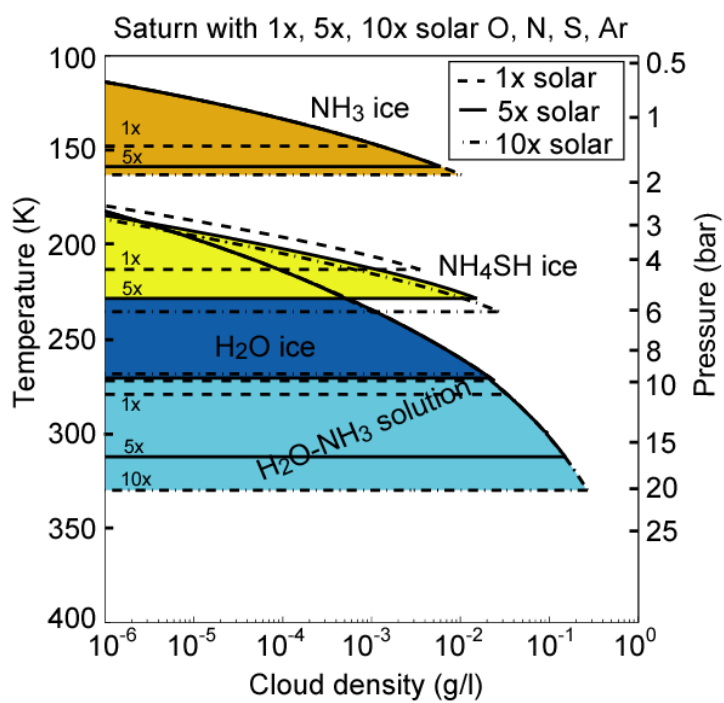

*Figure 2: Mean vertical distribution of cloud layers on Saturn, deduced from a simple thermochemical model. The cloud concentrations (in gram/liter) represent upper limits that are likely to be depleted by factors of 10-1000 due to precipitation and dynamics as on Earth (updated from Atreya and Wong (2005), using the Grevesse et al. (2005) solar elemental abundances).*

Ammonia, hydrogen sulfide and water vapor are all condensible species in Saturn's atmosphere. Their well-mixed atmosphere abundances yield, respectively, the elemental ratios N/H, S/H and O/H. On Saturn, $NH_3$ clouds are the topmost clouds, forming in the 1-2 bar region, depending on the enrichment of N, S, and O as $NH_3$ dissolves in water and forms a cloud of ammonium hydrosulfide upon combining with $H_2S$ (*Figure 2*). Thus, it would seem that ammonia is well-mixed below the 2 bar level, if thermodynamic equilibrium prevails and dynamics play no role in the atmosphere. It was evident from the Galileo probe measurements that the atmosphere of Jupiter is far from being ideal. Saturn is no different. The Cassini VIMS observations (Baines et al. 2009) show that a great degree of convective activity extends to at least several bars in Saturn's atmosphere as well. As a consequence, well mixed ammonia may only lie much deeper. In fact a true measure of well-mixed $NH_3$ could come only from the atmosphere below the water cloud since the $NH_3$ above these clouds may already be depleted due to solution in water. The well-mixed part for $H_2S$ is below the base of the $NH_4SH$ cloud or >6 bar level. Water forms the deepest cloud. Its base could range from 10 bars (ice cloud only, with 1x solar O/H) to >20 bars (an aqueous ammonia droplet cloud, with 10xsolar O/H). Again, because of convective processes, the well-mixed water may not be found until 50-100 bars. Thus water places the most severe constraint on the well mixed region of the atmosphere. Water is critical to determine, as it was presumably the original carrier of the heavy elements that formed the core of the planet and comprised 50-70% of its core mass according to formation models. The noble gases and all the isotopes listed above are accessible at pressures less than 10 bars, however, and are therefore an important design consideration for future probe missions to Saturn.

## 24.5 Saturn's atmospheric dynamics

As earlier chapters have abundantly demonstrated, the Cassini orbiter has provided a wealth of new data relating to the dynamics and circulation of Saturn's atmosphere, from the high stratosphere down to the deep troposphere, at least as far as the $NH_4SH$ cloud deck around 2-4 bars. This has led to major new insights into the chemistry and transport processes in the stratosphere, including the discovery of a substantial semi-annual oscillation of the zonal flow in the tropics. This is akin to the Earth's Quasi-Biennial and Semi-Annual Oscillations and is presumably driven by upward-propagating waves from the troposphere that break and dissipate in the middle stratosphere. The structure of the stratosphere is also seen to be strongly affected by seasonal variations, with substantial differences in temperature and composition between winter and summer hemispheres around solstice.

In the troposphere itself, new measurements from Cassini have discovered a surprisingly intense and compact vortex at the south pole, and have begun to measure how waves and eddies in Saturn's atmosphere interact to produce the well known banded pattern of zonal winds, clouds and hazes. Both Saturn and Jupiter appear to be in a special (`zonostrophic' – see Chapter 7) dynamical regime, in which energetic stirring of the atmosphere takes place on relatively small scales, dominated perhaps by baroclinic instabilities or intense but highly intermittent convective storms that may be driven, at least partly, by moist condensation effects. An upscale turbulent cascade then passes energy to ever larger scales, but becomes highly anisotropic due to dispersive wave propagation, the end-point of which is the energizing of intense but remarkably persistent zonal jets. In this regime, however, the effects of mechanical friction are very weak except at the largest scales, so jets may grow to sufficient intensity to become unstable on large scales, at least locally, leading to the production of large-scale waves and eddies that also often appear to be highly persistent. Such wave systems may include Saturn's unique and mysterious North Polar Hexagon and 'Ribbon wave' patterns, which seem to be remarkably stable and long-lived. However, the mechanisms by which these persistent circulation patterns sustain themselves remain somewhat mysterious.

Despite the considerable progress that has been, and continues to be made, by the Cassini-Huygens enterprise, however, it is already clear that a number of key questions confronting theoreticians and modelers will not be solved by the range of measurements enabled by the current mission. Such questions are likely to require new developments within the modeling community and, even more challenging, new kinds of measurements beyond those offered by Cassini, even given further mission extensions.

As discussed in Chapter 6 by Fouchet et al., measurements of composition in Saturn's stratosphere reveal anomalies that cannot be explained or modeled by local chemical changes or conventional 'eddy diffusion' parameterizations of transport. Large-scale transport across and between hemispheres is evidently more important than hitherto realized, and will require new approaches to chemical modeling in Saturn's stratosphere, most likely involving a full representation of chemical transport in 2D or even 3D models.

Cassini measurements of clouds and aerosols in Saturn's atmosphere (see Chapter 8; West et al.) have partly confirmed earlier ideas concerning the composition, nature and distribution of $NH_3$ and $NH_4SH$, although the lack of direct spectroscopic evidence for ammonia ice or $NH_4SH$ indicates that our knowledge of cloud microphysics on Saturn has some way to go. The interaction between microphysics, thermodynamics and atmospheric transport and circulation on a variety of scales should be a major objective of future work, both in terms of modeling and new observations post-Cassini. The large-scale distribution of clouds is critically dependent on the strength and form of the circulation in the upper troposphere on the scale of the banded zonal jets, and this continues to be quite uncertain. In this respect, also, the deep water abundance on Saturn is a key parameter that remains relatively poorly constrained on a global level.

On convective scales (~100 km or less), high resolution imagery from Cassini has shown a wealth of structure that is only just beginning to be explored – especially in the vicinity of the intense polar vortex. Current discussion of the polar vortex has tended to centre around supposed analogies with the core of terrestrial tropical cyclones, but this may be premature. The interaction between individual moist convection cells and a larger scale balanced vortex circulation is critical to the dynamics of terrestrial hurricanes, but whether this interaction operates in the same way in Saturn's polar vortex remains to be confirmed.

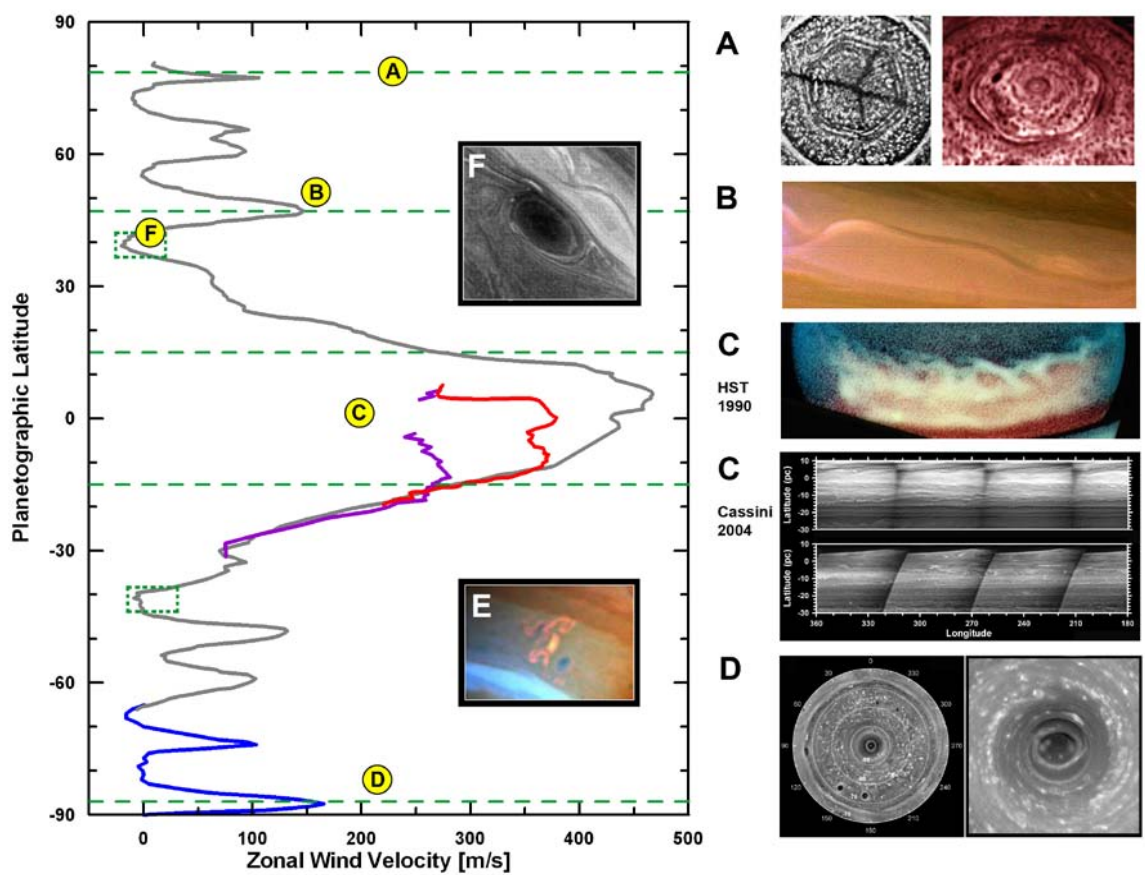

*Figure 3: General circulation of Saturn and relevant atmospheric features on its atmosphere (from Marty et al. 2008). Winds at cloud level, relative to the interior reference frame measured by Voyager, traced by the Voyagers (grey line) and Cassini data of the Southernmost latitudes (blue) and equatorial region in different filters (red and violet). Relevant meteorological structures appear on the insets: A) North polar hexagon in visible (Voyager 1) and infrared light (Cassini); B) The Ribbon; C) Saturn Great White Spot in the Equatorial Region in 1990 and the state of the Equator as seen by Cassini in the methane absorption band and continuum filters; D) The South Polar jet and the inner polar vortex; E) Convective storms seen by Cassini; F) Anticyclones from Voyager 1. The location of most convective storms appear marked with green dashed boxes.*

Much uncertainty in our understanding of the structure, dynamics and composition of Saturn's atmosphere stems from huge uncertainties in our knowledge of the nature and form of the atmospheric circulation beneath the ubiquitous upper cloud decks of $NH_3$, $NH_4SH$ and $H_2O$ condensates. Visible and near-infrared imagery only effectively senses the motions and structure of the uppermost $NH_4$ clouds in the main, although occasional glimpses of deeper seated structures are occasionally possible in relatively clear regions. Imaging in the thermal infrared around 5 µm wavelength by the VIMS instrument on Cassini has shown a wealth of structure on clouds at intermediate depths, most likely around 1-3 bars, largely representing the classical $NH_4SH$ cloud decks (see *Figure 4* below for an example). These structures appear quite differently and on a smaller scale than those apparent in visible images, suggesting a quite complex, baroclinic character for certain types of wave and eddies. Cassini has also begun to supplement the 5 µm images from VIMS with microwave remote sensing around 2 cm wavelength using the microwave RADAR instrument in passive mode. Microwave observations can penetrate significantly below the visible clouds, from a few bars at wavelengths of ~1 cm to several tens of bars at ~50 cm (de Pater & Dickel 1991; Briggs & Sackett 1989). This approach, utilizing long wavelength microwave measurements for remote sensing, is being developed for the JUNO mission to Jupiter (Janssen et al. 2005) and could be readily adapted for Saturn in due course.

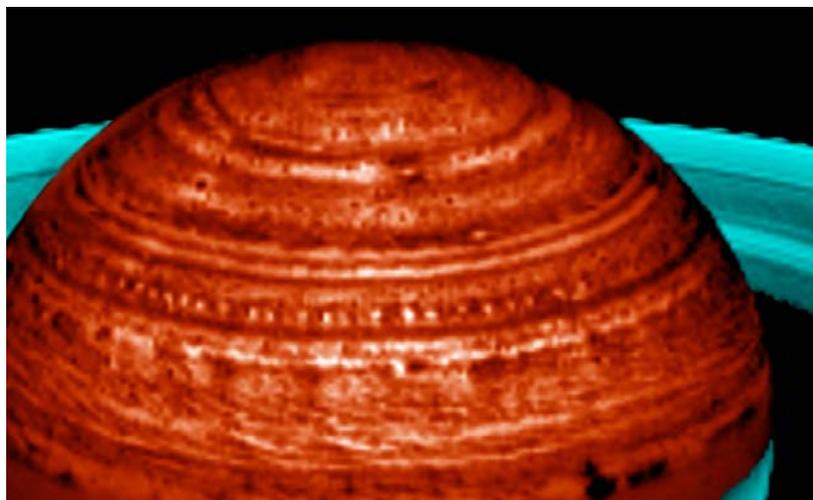

*Figure 4: Infrared image taken by the Cassini VIMS instrument in the 5 µm band of the northern hemisphere of Saturn, showing zonal bands and various waves and eddies (such as the `string of pearls' around 40$^o$N), representing features located in the vertical around pressure levels of 1-3 bars.*

But the issue remains that no remote sensing method, either from orbit or from the ground, is able to penetrate to levels deeper than a few tens of bars into Saturn's interior. This means that we are highly dependent on models to infer what may be happening beneath the clouds. Recent models (reviewed in Chapter 7 by Del Genio et al.), in particular, have highlighted a need to distinguish between the pattern of motions that may (or may not) be present to great depths below the visible clouds and the processes that may be driving them. It now seems clear (Lian & Showman 2008) that a deeply-penetrating meridional circulation pattern could be driven by energetic processes that are relatively shallow (such as moisture-enhanced convection in the water cloud layer), so a deep-seated circulation may not necessarily imply deep forcing. Moreover, where a circulation pattern penetrates very deeply into Saturn's troposphere, the electrical conductivity of the deep atmosphere may become a factor affecting the forces acting on fluid elements, in relation to Saturn's magnetic field. Theoretical and model studies of this magnetohydrodynamic aspect of Saturn's hypothetical deep atmospheric circulation are still in their infancy, but it is clear that Cassini alone will not be able to unravel them. Not least amongst these issues is the problem of Saturn's interior bulk rotation, whose period remains frustratingly illusive although some hints of a way forward have begun to appear (see Chapters 5, 7 and 9). This is because the radio/SKR measurements from Voyager and Cassini have proved to provide only an ambiguous and uncertain estimate of magnetic field rotation. Anderson & Schubert (2007), for example, have recently obtained a very different value for the interior rotation rate, based on a combination of gravity and radio-occultation measurements, with a period of 10 hours, 32 minutes, and 35 ± 13 seconds, though this is still regarded as somewhat controversial. Within a reference frame with such a rotation period, however, Saturn's zonal winds would appear more symmetrical about zero, with eastward and westward jets (other than the equatorial jet) of more or less equal strength, much as found on Jupiter.

For the future it will be necessary to focus on a set of key questions that should guide the design and objectives of both future spacecraft missions and theoretical and numerical modelling activities. In the dynamics and circulation arena these questions should include the following:

-What is the role of dynamical transport in determining the distribution of relatively short-lived chemical tracers in the troposphere and lower stratosphere?

-How do dynamics and microphysics interact to produce the observed clouds on Saturn (and Jupiter)?

-How is the cloud-level circulation maintained? What is the nature (and location) of both energetic forcing (moist convection, baroclinic instability...?) and large-scale energy dissipation?

-What is the nature and role of observed convective storm systems, coherent waves and stable vortices in the general circulation?

-How deep do the observed cloud-level circulation systems penetrate? Do they reach levels where electrical conductivity becomes important? If so, how does this affect the nature of the deep circulation?

-What is the nature of the polar vortices on Saturn (and the other gas giant planets)? How are they sustained? What determines their stability and what is their role in the planet-scale circulation?

This is not intended to be an exhaustive list, and doubtless others working in the field will come up with additional issues that are felt to be important and compelling. But these are a core of recurring questions that have challenged the planetary atmospheres community for decades now, and will continue to hold up further progress in understanding more detailed aspects of Saturn until they are resolved. This will require observations of Saturn's atmosphere, in particular at deep levels, but it also will require ambitious hydrodynamic modeling of the Saturn's atmosphere and interior including all relevant physical processes (see the chapter by Del Genio et al.).

## 24.6  Saturn's magnetosphere

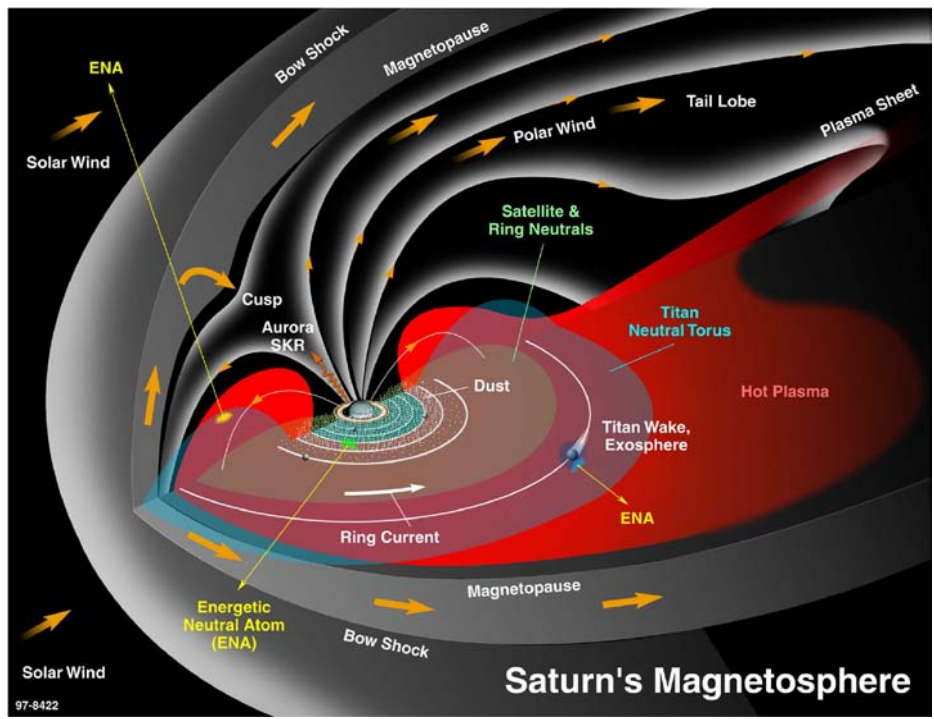

*Figure 5: An overview schematic of Saturn's magnetosphere revealing the complex nature of physical processes therein (from Krimigis et al. 2004).*

An initial understanding of Saturn's magnetosphere was obtained by three spacecraft flybys in the late seventies and early eighties, that of Pioneer 11 and Voyagers 1 and 2. These limited in-situ observations confirmed via plasma, fields and particles data that Saturn's magnetosphere resulted from the interaction of the solar wind plasma with Saturn's internal planetary magnetic field trapping plasma and energetic charged particles. The magnetosphere seemed to share many of the characteristics of the Earth's solar wind dominated magnetosphere and Jupiter's rotation-dominated one (Dougherty *et al.*, 2004). However it was clear that this magnetosphere is unique as a result of the extremely diverse nature of the coupling mechanisms which exist between the numerous components which made up the Saturn system, including the solar wind, the ionosphere of Saturn; Titan, the icy satellites, the rings and dust and neutral clouds. Addressing the importance and complexity of these various mechanisms was one of the main science objectives of the Cassini-Huygens mission (Blanc *et al.*, 2004) with a major strength of an orbiting mission such as Cassini being the potential to be able to resolve temporal from spatial processes via extended coverage of radial distance, latitude, longitude and Saturn local time.

The primary magnetospheric science objectives for the first four years of the Saturn tour by Cassini can be described as follows:
- To characterize the magnetic field, plasma and energetic particle population within the magnetosphere as a function of time and position
- To determine the relationship of the magnetic field orientation to Saturn's kilometric radiation
- Investigate Titan-magnetosphere interactions (this topic is described in the companion Titan book)
- Study the interaction between the magnetosphere and the icy satellites
- Investigate the interaction of the rings and the magnetosphere

On completion of the primary Cassini mission some major discoveries have resulted which will be briefly described (further details are given in chapters 10 - 13). Resolution of some of the objectives is still incomplete and requires further observations during the Cassini Extended Mission (XM) as well as during the planned extension onto 2017, known as the Solstice Mission. Following on from the discoveries at Titan and Enceladus in particular there is the possibility of a future Titan orbiter with numerous Enceladus flybys which will allow further study of the Saturn environment. Recent studies of a Saturn probe mission have also been carried out between US and European scientists (Marty. et al., 2009), with the primary goal of determining the abundance of the heavy elements, as was done at Jupiter with the Galileo probe in 1995. Complementary remote sensing observations from such a mission have the potential of yielding other valuable pieces of information including that on the core and the magnetospheric environment of the planet.

Characterization of the magnetosphere as a function of time and position has been carried out by a survey of the inner magnetosphere via the MAPS instruments onboard Cassini. Some major discoveries include: (i) That solar wind control is relatively weak compared to the rotational and mass-loading effects ; (ii) A new radiation belt has been found inside of the D-ring; (iii) Imaging of the rotating dynamic ring current; (iv) The magnetosphere has a time-varying rotation period. However complete characterization is incomplete since the orbit of Cassini did not reach the magnetotail reconnection region and the neutral sheet during the prime mission. Observations during the XM and SM will fill this gap in local time and allow a more complete characterization of the magnetospheric processes. In addition the first in-situ observations of Saturn's auroral zones were only achieved late in the prime mission (see Ch. 13 and PSS special issue articles reference therein) with further orbits covering this region in the XM and SM. This data is critical in order for a better understanding to be gained of the physical processes driving the aurora. In order to best understand a complex three-dimensional system such as the magnetosphere as much temporal and spatial coverage as possible is desirable in order to separate the importance of the different plasma processes arising. We have at the end of the prime mission a 4-year timeline allowing a much better understanding of the effect of temporal variations; however extended mission observations will further refine this understanding as well as take account of the seasonal effects which arise at Saturn. Higher order moments of Saturn's internal magnetic field have finally been resolved during the prime mission (Burton *et al.*, 2008) however in order to accurately determine the planetary field and better resolve the effect of the ring ionosphere on observations inside of the D-ring we require much more complete spatial coverage at a wide range of latitude, longitude and radial distances.

Our understanding of the relationship between Saturn's magnetic field orientation and the SKR is much advanced following the prime mission observations. It has been determined that the SKR period does not represent the internal rotation period of the planet, that the SKR period is variable and continues to evolve and that many magnetospheric phenomena have a period similar to the SKR period (Ch. 11) However there are still unresolved questions which require further measurements during the XM; the prime mission orbit in fact only resulted in very limited observation in the 3-5 Rs region, the 4-year mission to date has not allowed for a long enough observation in order to be able to distinguish between competing theories of why the SKR period varies and why it does not reflect the internal rotation period of the planet; and we are still yet to determine what this internal rotation period is.

Many major icy satellite discoveries have been made (Chapters 19-22) here we will focus on the interactions of the icy satellites and the magnetosphere. Some major discoveries include the discovery of a dynamic and exotic atmosphere at Enceladus, which led to the discovery of the plumes of Enceladus and confirmation that this moon is the source of Saturn's E-ring (Chapter 22). In addition a unique charge particle interaction with Rhea may be due to cloud and dust particles trapped within the Hill sphere (Chapter 12). There are as yet many unexplored areas concerning satellite/magnetosphere interaction and the XM and SM will yield additional close flybys (both upstream and downstream) of Mimas, Dione, Thetys and Rhea; and further Enceladus flybys will allow details of the plumes to be further studied as well as it's interaction with the magnetosphere.

The magnetospheric interaction with Saturn's rings has become more complex than originally though due the discovery of ring ionospheres (Chapter 14) and this requires further study. Saturn Orbit Insertion (SOI) demonstrated the unique nature of the region just above the main ring system. However due to the critical nature of the SOI burn, this orbit was not configured for prime science observations and additional close periapses during the XM and in particular the end of mission scenario linked to the SM which will consist of numerous polar orbits inside of the D ring will enable a clear determination of the ring ionosphere and its properties to be made as well as help constrain the internal planetary magnetic field.

## 24.7 Saturn's rings

Saturns rings are a perfect, and the closest, example of an astrophysical disk. Unlike protoplanetary disks, material cannot accrete inside rings because of the strong planet's tides. However, since tidal forces decrease with the distance from Saturn, at the ring's outer edge (A and F ring) accretion and erosion processes are expected to occur and are manifested by the formation of Jeans Toomre waves in the A ring ("wakes') or clumps in the F ring (Esposito et al., 2008). Taking advantage of a multi-scale approach, exploring rings would help to understand fundamental astrophysical processes (gravitational instabilities, gap-opening in disks, accretion, viscous evolution etc..) via remote sensing and, ideally, direct observation. Of course rings are also interesting for themselves, and are still poorly understood. Whereas the Cassini spacecraft has brought outstanding new data, some fundamental questions remain unanswered about its origin and evolution, and the milestone of future exploration would be in-situ images and spectra, with spatial resolution better than 10cm. On the basis of theoretical arguments and numerical simulations (see chapters by Schmidt et al. and Charnoz et al.) a lot of large scales processes and evolutionary processes (viscous spreading, gap opening, evolution of particle size distribution, accretion, meteoritic bombardment, brightness asymmetries) depends on the micro-scale structures (at scales <10 m) and 3-dimensional organization of the particle, which have been never observed. Stellar occultation data (Colwell et al., 2006, 2007, Eposito et al., 2008, Hedman et al., 2007) allowed probing the rings with resolution ~ 10m in 1 dimension. But the detected structures could be aggregates rather than individual particle. Among the outstanding question is the rings' mass. It is believed to be of the order of Mimas' mass, but could be much larger (Esposito et al., 2008, Charnoz et al., 2009, see also the chapter by Charnoz et al.). The mass is a determinant parameter for understanding the long term evolution of the rings. Another point is the composition of Saturn's rings: They seem to be made of almost pure ice with some contaminants (see e.g. Nicholson et al., 2008), defying any formation scenario. We no review these points in more details.

*Direct imaging of particle size distribution*

Are Saturn's rings young or old? What is their origin? Theses two questions are deeply linked, but a long-standing paradox arises when one try to answer them jointly: numerous arguments suggest that the rings are young (<$10^8$ years) because of fast evolutionary processes (erosion due to meteoroid bombardment, surface darkening, viscous spreading, see the chapter by Charnoz et al.). However, in the current state of our knowledge, it seems very improbable that they originated less than 1Ga ago, mainly because of the too low cometary flux (Harris, 1984, Charnoz et al., 2008, and chapter by Charnoz et al.). To solve this paradox, it is proposed that the ring material is constantly reprocessed due to self-gravity and collisions, and thus, may appear much younger. In order to constrain this "cosmic-recycling" process, a detailed knowledge of the size distribution would be an invaluable data. Indeed, different material strength, different surface sticking properties and different accretion regimes (gravity vs. surface-sticking) would lead to different size distributions. Voyager and Cassini radio occultation experiments allowed a rough estimate of the size distribution, in the 1cm-10m range (e.g. Marouf et al., 1983), but it is somewhat model dependant. The intermediate-size range (~100 m) has been probed indirectly by the detection of 100m-1km moonlets or aggregates (Tiscareno et al., 2006, Esposito et al., 2008). However, when put together, these measurements still do not draw a coherent picture because (1) we do not know if we are observing individual particles or aggregates and (2) these measurements are taken at different radial locations whereas we know that the size distribution should change with distance (Nicholson et al., 2008, see also chapters by Schmidt et al. and Charnoz et al.). Only direct imaging at different radial locations in the rings would allow to unveil all these degeneracies and thus provide new and strong constrains on the particle size's distribution and evolution.

Indeed accretion and erosion within the rings may strongly alter the particle's size distribution. Despite the strong tidal field of Saturn, limited accretion is theoretically possible (e.g. Canup and Esposito 1995, see chapter by Charnoz et al. and by Schmidt et al.). Ring particles have weak cohesive forces, and therefore can assemble into transient structures much larger than an individual ring particle. Thus, an exotic accretion physics takes place, resulting in the formation of temporary aggregates, either called "wakes" in the A rings, where material assemble into elongated structures (see chapter by Schmidt et al.) or further away, close to the A-ring's edge or in the F ring, the aggregates can accrete into moonlets called "ringmoons" (Esposito and Colwell 1993). The presence of 100m objects in the A ring has been revealed by the tidal arm they imprint in their surrounding (Tiscareno et al., 2008). Are theses objects primordial, fragments of larger bodies, or are they the natural product or local accretion? Direct imaging of "propellers" would be invaluable to understand the ring evolution and origin. It is also possible that moonlets embedded in the rings are made of two components: an outer shell of ring-particles accreted at the surface of a dense core that could be anterior to the formation of the ring (as was suggested by the odd shape of Pan and Atlas, see Porco et al., 2007, Charnoz et al., 2007).

*Microstructuration*
Only a handful of ring structures are explained, and it is thought that a substantial number of observable phenomena (brightness asymmetry in the A ring, fractal appearance, ring texture, sharp edges…, see chapter by Colwell et al.) could be the large scale manifestation of "microstructuration" : collective structuration processes occurring below the kilometre scale. For example, this could be either gravitational wakes in the A ring, viscous overstabilities in the B ring (see chapter by Schmidt et al.), self-organization of particles nearby a satellite resonance (Shepelyansky et al., 2009). Gravitational wakes in the A ring have been predicted for long (Toomre et al. 1964) and their wavelength is about 100 m. Direct observation of wakes would allow to constrain their morphology (size, orientation, separation etc…) the ring's surface density, viscosity and the Toomre Q parameter (although some of these parameters could be measured by indirect means, see e.g. Hedman et al., 2007). In addition, Daisaka et al. (2001) have shown that the presence of gravitational aggregates in Saturn rings result in a larger macroscopic viscosity than for a swarm of individual particles, which would have strong consequences on the rings lifetime. The B ring is maybe the most mysterious one due to its high density leading to non-linear effects. Overstability in the B ring produces parallels transient structures, with ~50m wavelength, that could explain the "microsillon" appearance of the B ring. Some stellar occultation data (Colwell et al., 2007) have provided further evidences that over-stability is still active in the B ring, but densest regions of the rings could not be investigated.

*Rings thickness*
The main-ring thickness has never been measured because the edge-on brightness of Saturn's rings is dominated by the dusty F ring, which is dynamically excited. We hence have no direct measurement of the A,B,C,D ring thickness, which is a critical dynamical parameter, like in any astrophysical disk. In particular, the age of the rings is directly linked to their thickness (Esposito et al., 1986) since the timescale of viscous spreading depends closely on the local particle velocity dispersion. Photometric arguments suggest $H<1$km (Nicholson et al. 1996), but dynamical arguments suggest values of the order of, or less than, 5m in the B ring (Salo 1995).

*Chemical composition: the mystery of silicates*
Saturn's rings are composed mainly of water ice (see e.g. Cuzzi & Estrada 1998, Poulet et al. 2003, Nicholson et al. 2005), and a few contaminant brought by the meteoroid bombardment. The abundance of silicates must be lower than 1% if uniformly distributed within the rings particles (Grossman 1990). Such purity is difficult to explain in any cosmogonic scenario of Saturn's rings, and there is no obvious mechanism that could remove silicates from Saturn's rings (see for example Harris 1984, Charnoz et al., 2008). However, when probing the material composition by remote-sensing, the penetration depth of light is about of the order of the wavelength, so that only the particle surface is probed and the bulk of the dense B ring remains maybe unobserved. Maybe some silicates may hide in the core of big ring particles, but we have no clue about this.

*Rings' mass*
The total rings' mass is still unknown. It is thought to be of the order of Mimas' mass (Esposito et al., 1983). However, there are strong suspicions that it could be much larger (see e.g. Stewart et al., 2007), which in turn would help solve some aspects of the question of the ring's lifetime and origin (see chapter by Charnoz et al.). A lot of mass could be stored in the B ring. Unfortunately, estimating its mass through remote sensing is difficult because it is opaque even to radio emissions (normal optical depth can be larger than 5) and its surface density shows strong spatial and temporal variability (Colwell et al., 2007). A direct measurement of the mass of Saturn's B ring may be obtained from a gravity field inversion after flybys around the entire ring system.

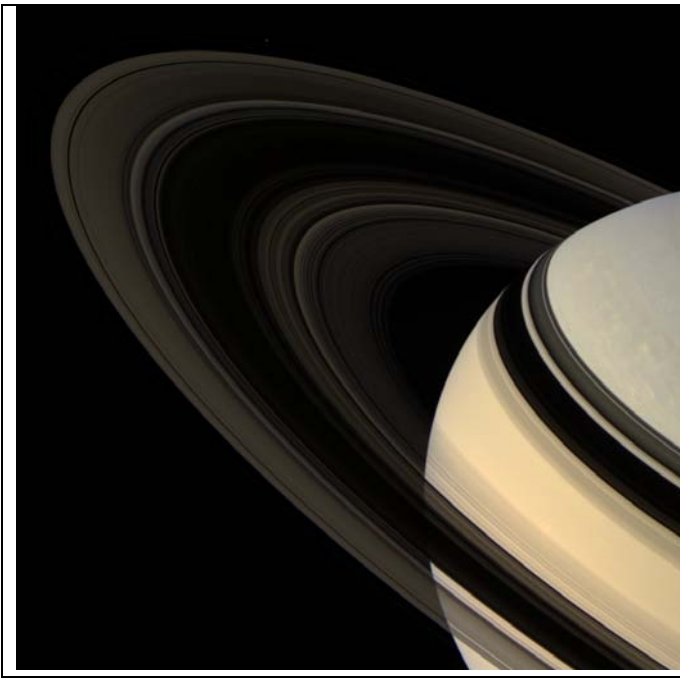

Figure 6-: *Saturn rings observed in transmitted light by the Cassini/ISS camera system. In this point-of-view, denser rings appear darker because they block the light coming from the Sun. The B ring appears as a wide dark lane in the middle of the ring system because of its high surface density.*

*Credits : NASA/JPL/Space Science Institute*

## 24.8 Saturn and the formation of the Solar System

Saturn formed about 4.56 Ga ago, probably within a few Ma of the formation of the Sun, and at an epoch when a circumstellar disk, mainly made of hydrogen and helium, still surrounded the infant Sun (e.g. Pollack et al. 1996, Wuchterl et al. 2000, Cassen 2006). On the other hand, the terrestrial planets were almost certainly fully formed later, in the 10-30 Ma or so after the gaseous disk had disappeared (e.g. Chambers 2004). Based on various observations like the structure of the Kuiper belt, the present orbital parameters of the planets in the Solar System and the traces of a late heavy bombardment in the inner system, it appears that Saturn (and the whole system of outer planets) was initially closer to the Sun, i.e. around 8 AU, instead of 9.54 AU now (Tsiganis et al. 2005). According to this scenario, known as the "Nice model", it is Saturn's crossing of the 1:2 resonance that has led to the so-called late heavy bombardment responsible for the formation of the lunar basins 4 Ga ago, or about 500 Ma after the formation of the first solids in the system (Gomes et al. 2005).

However, not much is known concerning the early phase (few Ma) when the gaseous protoplanetary disk was still present and the giant planets were forming. Models of planet formation that attempt to explain the observed ensemble of planets orbiting other stars (Ida & Lin 2004, Alibert et al. 2004, Thommes et al. 2008) indicate that giant planets such as Jupiter and Saturn grew from planetary embryos that had reached by accretion masses of a few times the mass of the Earth. However, it is not clear whether these protoplanets had migrated significantly (e.g. Alibert et al. 2005, see also Cresswell & Nelson 2008), or whether a nearly joint formation of Jupiter and Saturn locked them into a resonance with little migration (e.g. Masset and Snellgrove 2001, Morbidelli et al. 2008). Also we don't know whether Jupiter, Saturn, Uranus and Neptune formed at the same time or in sequence, and whether they formed in a protoplanetary disk still massive or relatively light. There are arguments that indicate that planetary embryos grew while the disk was being evaporated (see Ida, Morbidelli & Guillot 2008), so that it seems plausible that Jupiter would have formed first and thereby acquired the largest mass in hydrogen and helium while Saturn would have formed a little after. In that scenario, Uranus and Neptune would have formed near the end of the lifetime of the protosolar disk. This, however, remains a conjecture given the scarce amount of evidence and constraints. We need to better constrain the structure and precise composition of the giant planets in order to truly understand how the Solar System was formed.

In this mysterious early phase of the formation of the Solar System, two planets stand out, because they were probably the first to appear and had a tremendous impact on the structure of the planetary system that formed around the Sun: namely, Jupiter and Saturn. It is therefore crucial that their characteristics be compared with a similar level of detail. Jupiter's atmospheric composition has been measured by the Galileo probe, and Jupiter's interior will be further examined by the Juno mission. We presently lack similar observations for Saturn. As discussed in §24.2, we would want to compare Jupiter and Saturn's central core masses as well as their total mass of heavy elements to know how they formed, and possibly where. A striking example of this need for a comparison is tied to the abundance of noble gases measured in Jupiter by the Galileo probe, but not in Saturn.

In Jupiter, noble gases (Ar, Kr, Xe) are enriched compared to their abundance in the Sun by a factor ~2. This is puzzling and still unexplained, mostly because noble gases (particularly argon) are very difficult to trap into solids and deliver into the planet's atmosphere. Several explanations have been put forward: (i) Jupiter was formed at very low temperatures, at a place where even argon would be able to condense onto grains (Owen et al. 1999); (ii) Jupiter was impacted with planetesimals made of crystalline ice in which an efficient clathration process occurred (Gautier & Hersant 2005; Alibert et al. 2005); (iii) Jupiter's formation occurred late in a photoevaporating disk in which the midplane had been progressively enriched in all elements capable of sticking to grains in the disk's very cool outer regions, including noble gases (Guillot & Hueso 2006). These scenarios explain Jupiter's case, but yield very different answers at Saturn, as shown by *Figure 7*. Measuring Saturn's abundance in noble gases would allow a decision between these very different possibilities.

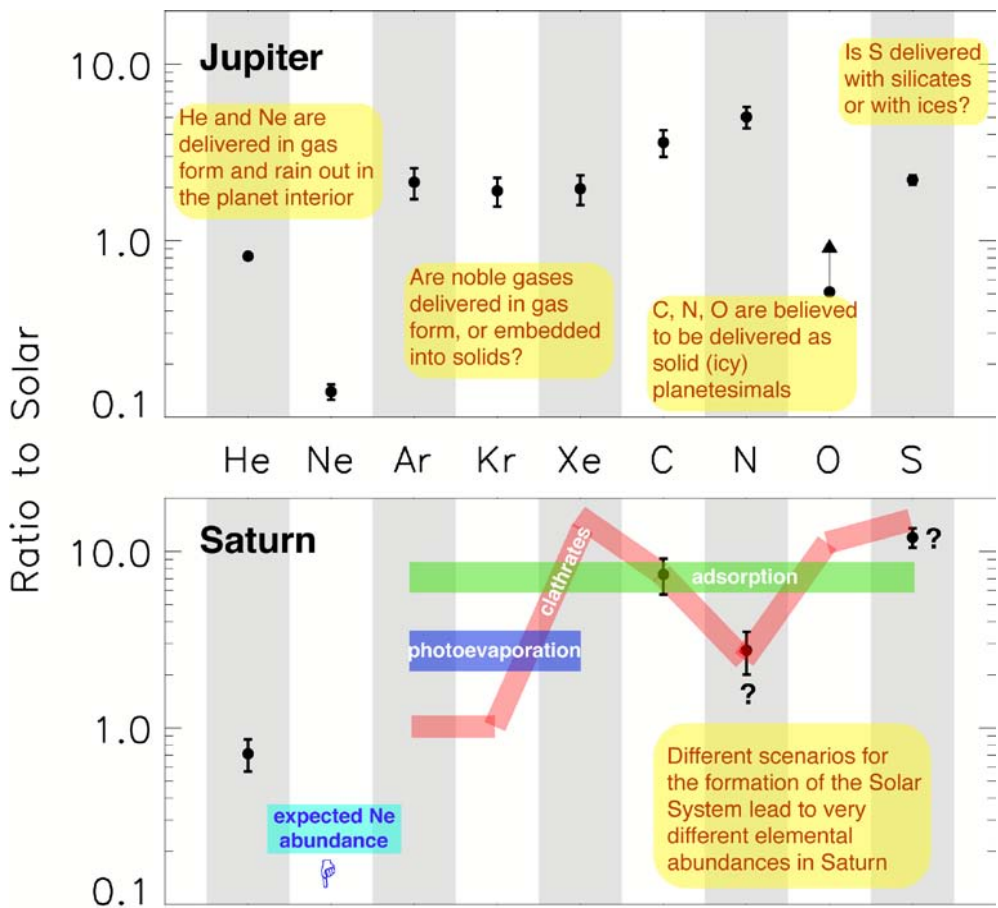

*Figure 7: Elemental abundances measured in the tropospheres of Jupiter (top) and Saturn (bottom) in units of their abundances in the protosolar disk. The elemental abundances for Jupiter are derived from the in situ measurements of the Galileo probe. The abundances for Saturn are spectroscopic determinations from Cassini for He/H and C/H, and model-dependant ground based measurements for N/H and S/H. A determination of the abundance of noble gases in Saturn would allow distinguishing between different formation scenarios whose predictions are shown as green, blue and pink curves, respectively (see text). [Adapted from Marty et al. (2008)].*

### 24.9   The means of Saturn's future exploration

In this chapter we have attempted to demonstrate that to satisfactorily address the questions of the formation of the outer solar system, at the very minimum a comparative study of the two gas giant planets, Jupiter and Saturn, is essential. While Jupiter has been explored with flybys (Pioneer, Voyager, Cassini), an orbiter (Galileo), a probe (Galileo), and will be further investigated with the Juno orbiter, the complete exploration of Saturn still has long way to go despite the highly successful Cassini orbiter mission. The bulk composition, particularly the abundance of heavy elements, existence and magnitude of the core, deep meteorology and dynamics and internal rotation period are some pieces of the Saturn's puzzle that need to be pursued vigorously in any future exploration of Saturn. Some questions can be addressed with a flyby spacecraft while others may require an orbiter. In either case, entry probes are key. Complementary observations from 1 AU also have an important role to play.

*24.9.1   Flyby*

A flyby of Saturn is among the simplest solution among space missions. Its scientific return from Saturn would be highly valuable with a properly chosen flyby geometry. With a very close flyby of a few thousand kilometers above Saturn's cloud tops at pericenter, it is possible to obtain an extremely accurate measurement of the planet's gravity field. This would allow to determine whether Saturn's winds extend deep into the interior. This determination and a precise measurement of the gravitational moments of the planet would allow to much better constrain the planet's interior structure. Furthermore, the flyby would yield Saturn's true magnetic field, unfiltered by the rings. As with

Juno at Jupiter (Janssen et al. 2005), microwave radiometry from the flyby spacecraft can allow the determination of $NH_3$ and $H_2O$ in the well-mixed atmosphere, hence the oxygen and nitrogen elemental abundances, that are crucial components of Saturn's bulk composition since water was presumably the original carrier of heavy elements. The determination of remaining heavy elements and isotopes would still require entry probes, however.

*24.9.2 Orbiter*
To date, most infrared remote sounding has been carried out using nadir methods which, although sensitive to trace constituents, are very limited in vertical resolution. Moreover, a significant part of Saturn's (and Jupiter's) middle atmosphere is still unexplored. This gap, the "ignorosphere", is important to fill, however, for a complete understanding of Saturn's physico-chemical workings. A valuable extension of presently used methods to include systematic limb-sounding could allow much more detailed coverage of the stratosphere, both for dynamics and composition, especially if carried out systematically from a stable, near-circular polar orbiting platform for a significant fraction of a Saturnian year. This extension of the record of horizontal velocity fields from cloud tracking in the visible and infrared would also make a valuable contribution to our understanding of the turbulent nature of Saturn's atmosphere. Again, this would be particularly valuable if carried out systematically and globally, over a long period of time from a suitable orbiting platform, and at higher spatial resolution even than achieved by Cassini in order to resolve individual convective circulations. Even though such measurements would be restricted in altitude coverage to pressure levels less than 1-2 bars, this is probably the only feasible means in the foreseeable future to obtain the kind of detailed, global information on winds and transport of energy, momentum and vorticity necessary for comparison with the most sophisticated numerical models of atmospheric circulation.

At deeper levels, however, the only means of recovering information on bulk motions in the planetary interior would have to come from highly detailed and accurate measurements of the gravity and magnetic fields, as suggested above for a flyby. Following the example of Juno, this ideally requires the placing of a suitably instrumented spacecraft into a very low altitude polar orbit with the aim of measuring the high order gravitational moments > $J_{10}$. As noted by Hubbard (1999), were Jupiter's (or Saturn's) cloud-level banded zonal flow to penetrate barotropically into the deep interior, the need for the pressure field to accommodate a deep geostrophic flow would require adjustments in the interior mass distribution that would manifest themselves in the fine structure of the external gravity field. Because such high order moments decay rapidly in amplitude with distance from the centre of the planet, it is necessary to measure them at very low altitude – in Saturn's case within the inner radius of its ring system. Under current plans, NASA's Juno mission will achieve this objective for Jupiter, and a similar approach will also be necessary to constrain Saturn's deep zonal flow. To complete the picture, if Saturn's banded zonal flows were to penetrate to the deep interior, it is presumably likely that they would perturb the magnetic field on scales similar to the gravity field. Thus, high precision measurements of the high order moments of the magnetic field would be expected to provide important constraints on patterns of deep-seated circulations.

Microwave radiometric measurements done from an orbiter, as opposed to a flyby, will also have the advantage of providing a map of deep $NH_3$ and $H_2O$ over Saturn, as at Jupiter with Juno. This would be particularly useful for understanding the convective processes in the deep atmosphere that could prevent these species from being well-mixed in localized regions even at very deep levels of Saturn's atmosphere.

*24.9.3 Probes*
Entry probes offer the potential advantages of high vertical resolution in the measurement of vertical profiles of temperature, winds, clouds and composition during their descent. They are also capable of measuring the much needed bulk composition and isotopes, including the noble gases, He, Ne, Ar, Kr and Xe, together with their isotope abundances, condensable species, $H_2O$, $NH_3$ and $H_2S$, $CH_4$, isotopic ratios of 14N/15N, 3He/4He, D/H. The well-mixed abundances of only $CH_4$ (infrared) and $NH_3$ and $H_2O$ (microwave radiometry from flyby or orbiter) can be measured by remote sensing. Probes are essential for the other measurements in the list. However, as the Galileo probe demonstrated, the condensable species ($H_2O$, $NH_3$ and $H_2S$) may be vulnerable to being unrepresentative of the planet as a whole unless a significant number of locations are sampled. The practicalities of communication with such probes also places limits on the depth from which they can recover information, unless they can be made sufficiently sophisticated to adjust their buoyancy autonomously and resurface to relay the results of their measurements at a later time. But even relatively modest probes could provide much-needed information on the profiles of static stability and humidity that would significantly constrain current models of moist convection on Saturn, especially if this can be recovered from depths greater than 10 bars. Profiles of the concentrations of condensable species would be particularly valuable, especially if a representative range of meteorological phenomena (e.g. belts, zones, plumes and hot-spots) could be sampled. Doppler wind measurements at a number of locations would also allow extension of the cloud-top wind patterns to greater depths, which may be diagnostic of processes maintaining the banded system of jets and clouds.

A probe-orbiter or probe-flyby combination would provide the most crucial information necessary to unravel Saturn's mysteries, as the noble gases and the isotopes can be sampled at relatively shallow depths whereas deep water and ammonia can be mapped with microwave radiometry from an orbiter. The orbiter also has the potential of providing data for discerning the presence of a core.

*24.9.4   Microprobe in Saturn's rings*
In order to achieve the rings science objectives listed above, the most valuable project would be an in-situ mission into Saturn's rings. Ideally, a big probe with multiple sensors should be dropped into the rings. However, the technical difficulties to inject the spacecraft into the rings (bring it close to Saturn, brake it to a nearly circular orbit, keep the instruments safe etc…) and to navigate it (requiring an efficient propulsion system to keep the spacecraft above the ring plane) imply that this is a project for the long-term future. Fortunately, a large fraction of the science objectives could be achieved with a couple of instruments (a camera and/or a spectro-imager) and a communication system onboard very simple probes, called "microprobes". A spacecraft would drop microprobes at different locations into the rings (A,B,C by order of priority) on impact trajectories. On close approach this would provide opportunity for very high resolution images of ring particles. A difficulty would be to put the microprobes on very low inclination trajectories to maximize the time for science return before impacting into the ring material. In addition, if a gap is targeted, and if a probe survives the ring plane crossing, this would allow direct measurement of ring's thickness. Data would have to be collected, compressed and transmitted in real time, either to a spacecraft or directly transmitted to Earth, as was shown to be possible by the Huygens probe from Titan's surface.

*24.9.5   Observations from 1AU*
Earth-based planetary astronomy has always played an important supporting role in the exploration of the planets and Saturn is no exception. A long-term database is important to discern regular and irregular temporal changes, and is practical only with observations carried out from the earth. Observations in the infrared with ground-based telescopes can provide valuable data on certain stratospheric molecules and the tropospheric meteorology particularly above the ammonia cloud tops. High-resolution submillimeter measurements with the ALMA telescope have the potential of extending the coverage. Space-borne telescopes such as Herschel and the James Webb Space Telescope will undoubtedly open new vistas into Saturn's atmosphere not accessible to previous ground-based or Earth-space observations. It should be stressed, however, the observations made from Earth are only complementary to, not a replacement for, the above probe-flyby/orbiter observations.

**24.10 Conclusions**

Saturn is a beautiful, intriguing, complex planet. Years of research, observations, experiments, theories, several flybys, measurements in orbit by the Cassini-Huygens spacecraft, all have greatly expanded our knowledge of this giant planet. They have at the same time highlighted or raised many questions that remain unanswered: What is Saturn's interior composition? How does the planet evolve? Why can't we predict its storms? What maintains its strange magnetic field? What is the origin of its rings? How was the planet formed?

A continued exploration of Saturn is essential to understand our Solar System and progress in areas of science as diverse as the study of atmospheric dynamics to dynamo theory and the formation of planetary systems. Future exploration of Saturn will benefit from an appropriately parallel approach to Jupiter's in the past, i.e. with a complement of similar instruments and techniques including in situ measurements. The comparison of similarities and differences will yield a gain of knowledge much greater than the sum of the independent pieces of information obtained for either planet. Even very different theories of the formation of the Solar System could not be tested by measurements in only one planet; they require comparison of key data at at least the two gas giant planets.

Saturn's exploration should proceed with a variety of methods. Sending a probe into its atmosphere, even at moderate pressures of a few bars will allow for a unique comparison with the Galileo measurements at Jupiter, including data on the heavy elements, clouds and winds. Measurements of the planet's deep water abundance, interior rotation, magnetic field inside the D ring are extremely important both for understanding the planet's interior, evolution, origin, meteorology and magnetic field. They require either a flyby or a very close orbiter. Similarly, important gains in our understanding of planetary meteorology and dynamics can be made by monitoring Saturn's clouds and weather systems at very high spatial resolution and, ideally, over relatively long timescales. Continued monitoring of the planet remotely with ground- and space-based instruments will be highly valuable, as they will provide complementary and synoptic information. Finally, the study of the rings itself also requires high spatial resolution images to detail its components, something that may be most easily obtained with microprobes.

## References

Alibert, Y., Mordasini, C., Benz, W. Migration and giant planet formation. A&A 417, L25-L28 (2004)

Alibert, Y., Mousis, O., Benz, W.: On the volatile enrichments and composition of Jupiter. Astrophys. J. 622, L145–L148 (2005)

Anderson, J. D., Schubert, G.: Saturn's gravitational field, internal rotation, and interior structure, Science, 317, 1384-1387 (2007).

Atreya, S.K., Wong, A.S.: Coupled chemistry and clouds of the giant planets – A Case for multiprobes, in Outer Planets (eds. Encrenaz, T., et al.), pp121-136, Springer-Verlag (2005). Also in, Space Sci. Rev., 116 (nos. 1-2), 121-136 (2005)

Atreya, S.K., Wong, M.H., Owen, T.C., Mahaffy, P.R., Niemann, H.B., de Pater, I., Drossart, P., Encrenaz, Th., A comparison of the atmospheres of Jupiter and Saturn: Deep atmospheric composition, cloud structure, vertical mixing, and origin. Planet. Space Sci. 47, 1243-1262 (1999)

Atreya, S.K.: Atmospheres and Ionospheres of the Outer Planets and their Satellites, Chapter 3. Springer-Verlag, New York (1986)

Baines, K. H., Momary, T. W., Fletcher, L. N., Showman, A. P., Roos-Serote, M., Brown, R. H., Buratti, B. J., Clark, R. N. Saturn's north polar cyclone and hexagon at depth revealed by Cassini/VIMS. Planetary and Space Sci., submitted (2009)

Bakos, G.A., et al. HAT-P-1b: A Large-Radius, Low-Density Exoplanet Transiting One Member of a Stellar Binary. Astrophysical Journal 656, 552-559 (2007)

Barbara J.M., Esposito L.W.: Moonlet collisions and the effects of tidally modified accretion in Saturn's Fring. Icarus **160**, 161–171 (2002)

Blanc M., S. Bolton, J. Bradley, M. Burton, T. E. Cravens, I. Dandouras, M. K. Dougherty, M. C. Festou, J. Feynman, R. E. Johnson, T. G. Gombosi, W. S. Kurth, P. C. Liewer, B. H. Mauk, S. Maurice, D. Mitchell, F. M. Neubauer, J. D. Richardson, D. E. Shemansky, E. C. Sittler, B. T. Tsurutani, Ph. Zarka, L. W. Esposito, E. Gruen, D. A. Gurnett, A. J. Kliore, S. M. Krimigis, D. Southwood, J. H. Waite, and D. T. Young, Magnetospheric and Plasma Science with Cassini-Huygens,*SpaceSci.Rev., 104*, 253-346, 2004

Briggs, F. H., Sackett, P. D.: Radio observations of Saturn as a probe of its atmosphere and cloud structure. Icarus 80, 77-103 (1989)

Cassen P.: Protostellar Disks and Planet Formation, In Extrasolar Planets, Saas-Fee adv courses vol 31 (eds Mayor et al.), pp 369-448 (2006)

Chambers, J. E.: Planetary accretion in the inner Solar System. Earth and Planetary Science Letters 223, 241-252 (2004)

Charbonneau, D., Brown, T.M., Burrows, A., Laughlin, G. When Extrasolar Planets Transit Their Parent Stars. In Protostars and Planets V, 701-716 (2007)

Charnoz S., Brahic A., Thomas P.C., Porco C.C. The equatorial ridges of Pan & Atlas: terminal accretionary ornaments ? Science 318, 1622 (2007)

Charnoz S., Morbidelli A., Dones L., Salmon J.: Did Saturn rings formed during the Late Heavy Bombardment ? Icarus 199, 413-428 (2009)

Colwell, J. E.; Esposito, L. W.; Sremčević, M.: Self-gravity wakes in Saturn's A ring measured by stellar occultations from Cassini. GeoRL **33**, L07201 (2006)

Colwell J. E., Esposito L. W., Sremčević M., Stewart G. R., McClintock W. E.: Self-gravity wakes and radial structure of Saturn's B ring. Icarus **190**,127-144 (2007)

Conrath, B., Gautier, D.: Saturn helium abundance: A reanalysis of the Voyager measurements. Icarus 144, 124-134 (2000)

Cresswell, P., Nelson, R.P. : Three-dimensional simulations of multiple protoplanets embedded in a protostellar disc. A&A 482, 677-690 (2008)

Cuzzi, J.N.; Estrada, P.R. : Compositional evolution of Saturn's rings due to meteoroid bombardment. *Icarus* **132**, 1-35. (1998)de Pater, I. & Dickel, R. (1991) Multifrequency Radio Observations of Saturn at Ring Inclination Angles between 5 and 26 Degrees, *Icarus*, **94**, 474-492.

Daisaka H., Tanaka H., Ida S.: Viscosity in a Dense Planetary Ring with Self-Gravitating Particles. Icarus 154, 296-312 (2001)

Dougherty, M. K., Kellock, S., Southwood, D. J., Balogh, A., Smith, E. J., Tsurutani, B. T., Gerlach, B., Glassmeier, K.-H., Gleim, F., Russell, C. T., Erdos, G, Neubauer, F. M. and Cowley, S. W. H., The Cassini magnetic field investigation, *Space Sci. Rev.*, **114**, 331-383, 2004

Esposito L.W., O'Callaghan M., West R.A..The structure of Saturn's rings: Implications from the Voyager stellar occultation. *Icarus* **56**, 439-452. (1983)

Esposito L.W., Meinke B.K., Colwell J.E., Nicholson P.D., Hedman M.H: Moonlets and clumps in Saturn's F ring. Icarus **194**, 278-289 (2008)

Flasar, M., and the CIRS Team: Temperature, winds, and composition in the Saturnian system. Science 307, 1247-1251 (2005)

Fletcher, L. N., Irwin, P. G. J., Teanby, N. A., Orton, G. S., Parrish, P. D., Calcutt, S. B., Bowles, N., Kok, R. de, Howett, C., Taylor, F. W.: The meridional phosphine distribution in Saturn's upper troposphere from Cassini/CIRS observations. Icarus 188, 72-88 (2007)

Fortney, J.J., Hubbard, W.B.: Phase separation in giant planets: inhomogeneous evolution of Saturn. Icarus 164, 228-243 (2003)

Fortney, J.J., Hubbard, W.B.: Effects of Helium Phase Separation on the Evolution of Extrasolar Giant Planets. Astrophysical Journal 608, 1039-1049 (2004)

Gautier, D., Hersant, F.: Formation and composition of planetesimals—Trapping volatiles by clathration. Space Sci. Rev. Space Sci. Rev. 116, 25–52 (2005)

Gomes, R., Levison, H. F., Tsiganis, K., Morbidelli, A. : Origin of the cataclysmic Late Heavy Bombardment period of the terrestrial planets. Nature 435, 466-469 (2005)

Grevesse, N, Asplund, M., Sauval, J.: The new solar composition. In: Alecian, G., Richard, O., Vauclair, S. (eds.) Element Stratification in Stars: 40 Years of Atomic Diffusion, pp. 21-30. EAS Publications Series (2005)

Grossman, A.W.: Microwave imaging of Saturns deep atmosphere and rings. Ph.D. Thesis, California Institute of Technology (1990)

Guillot, T. : The Interiors of Giant Planets: Models and Outstanding Questions. Annual Review of Earth and Planetary Sciences 33, 493-530 (2005)

Guillot, T. : The composition of transiting giant extrasolar planets. Physica Scripta Volume T 130, 014023 (2008)

Guillot, T., Hueso, R.: The composition of Jupiter: sign of a (relatively) late formation in a chemically evolved protosolar disc. Mon. Not. R. Astron. Soc. 367, L47–L51 (2006)

Harris A.: The origin and evolution of planetary rings. In *Planetary Rings*, R. Greenberg, A. Brahic, Eds., (Univ. Arizona Press, 1984) pp. 641- 659. (1984)

Hedman, M.M., Burns, J.A., Tiscareno, M.S., Porco, C.C., Jones, G.H., Roussos, E., Krupp, N., Paranicas, C., Kempf, S.: The source of Saturn's G ring. Science **317**, 653-656 (2007)

Hubbard, W.B.: Gravitational signature of Jupiter's deep zonal flows. Icarus 137, 357–359 (1999)

Hubbard, W. B.; Guillot, T.; Marley, M. S.; Burrows, A.; Lunine, J. I.; Saumon, D. S. Comparative evolution of Jupiter and Saturn. Plan. Space Sci. 47, 1175-1182 (1999)

Ida, S., Lin, D.N.C.: Toward a Deterministic Model of Planetary Formation. I. A Desert in the Mass and Semimajor Axis Distributions of Extrasolar Planets. ApJ 604, 388-413 (2004)

Ida, S., Guillot, T., Morbidelli, A.: Accretion and destruction of planetesimals in turbulent disks. ApJ 686, 1292-1301 (2008).

Ikoma, M., Guillot, T., Genda, H., Tanigawa, T., Ida, S. : On the Origin of HD 149026b. Astrophysical Journal 650, 1150-1159 (2006)

Janssen, M. A., Hofstadter, M.D., Gulkis, S., Ingersoll, A. P., Allison, M., Bolton, S. J., Levin, S.M. & Kamp, L.W. (2005) Microwave remote sensing of Jupiter's atmosphere from an orbiting spacecraft, Icarus, 173, 447-453.

Krimigis, S. M., D. G. Mitchell, D. C. Hamilton, S. Livi, J. Dandouras, S. Jaskulek, T. P. Armstrong, A. F. Cheng, G. Gloeckler, K. C. Hsieh, W.-H. Ip, E. P. Keath, E. Kirsch, N. Krupp, L. J. Lanzerotti, B. H. Mauk, R. W. McEntire, E. C. Roelof, B. E. Tossman, B. Wilken and D. J. Williams. (2004) Magnetosphere Imaging Instrument (MIMI) on the Cassini Mission to Saturn/Titan, *Space Sci. Rev., 114(1-4),* 233-329

Lian, Y. & Showman, A. P. (2007) Deep jets on gas-giant planets, Icarus, 194, 597-615.

Marouf, E. A.; Tyler, G. L.; Zebker, H. A.; Simpson, R. A.; Eshleman, V. R. : Particle size distributions in Saturn's rings from Voyager 1 radio occultation. Icarus **54**, 189-211(1983)

Marty, B., T. Guillot, A. Coustenis & the Kronos consortium: N. Achilleos, Y. Alibert, S. Asmar, D. Atkinson & S. Atreya, G. Babasides, K. Baines, T. Balint, D. Banfield, S. Barber, B. Bézard, G. L. Bjoraker, M. Blanc, S. Bolton, N. Chanover, S. Charnoz, E. Chassefière, J. E. Colwell, E. Deangelis, M. Dougherty, P. Drossart, F. M. Flasar, T. Fouchet, R. Frampton, I. Franchi, D. Gautier, L. Gurvits, R. Hueso, B. Kazeminejad, T. Krimigis, A. Jambon, G. Jones, Y. Langevin, M. Leese, E. Lellouch, J. Lunine, A. Milillo, P. Mahaffy, B. Mauk, A. Morse, M. Moreira, X. Moussas, C. Murray, I. Mueller-Wodarg, T. C. Owen, S. Pogrebenko, R. Prangé, P. Read, A. Sanchez-Lavega, P. Sarda, D. Stam, G. Tinetti, P. Zarka & J. Zarnecki (2008) Kronos: exploring the depths of Saturn with probes and remote sensing through an international mission, *Exp Astron.* DOI 10.1007/s10686-008-9094-9.

Masset, F., Snellgrove, M. : Reversing type II migration: resonance trapping of a lighter giant protoplanet. MNRAS 320, L55-L59 (2001)

Morbidelli, A., Crida, A., Masset, F., Nelson, R.P. : Building giant-planet cores at a planet trap. A&A 478, 929-937 (2008)

Nicholson, P.D., Showalter, M.R., Dones, L., French, R.G., Larson, S.M., Lissauer, J.J., McGhee, C.A., Seitzer, P., Sicardy, B., Danielson, G.E.: Observations of Saturn's ring-plane crossing in August and November 1995. Science **272**, 509-515 (1996)

Nicholson P.D., Hedman M.M., Clark R.N., Showalter M.R., Cruikshank D.P., Cuzzi J.N.,

Filacchione G., Capaccioni F., Cerroni P., Hansen G.B., Sicardy B., Drossart P., Brown R.H.,

Buratti B.J., Baines K.H., Coradini A.: A close look at Saturn's rings with Cassini VIMS. Icarus **193**, 182-212 (2008)

Owen, T., et al.: A low-temperature origin for the planetesimals that formed Jupiter. Nature 402, 269–270 (1999)

Pollack, J.B., Hubickyj, O., Bodenheimer, P., Lissauer, J.J., Podolak, M., Greenzweig, Y. : Formation of the Giant Planets by Concurrent Accretion of Solids and Gas. Icarus 124, 62-85 (1996)

Porco C. C., Thomas P. C., Weiss J. W., Richardson D. C.: Saturn's Small Inner Satellites: Clues to Their Origins. Science 318, 1602 (2007)

Poulet F., Cruikshank D.P., Cuzzi J.N., Roush T.L., French R.G.: Composition of Saturn's rings A, B, and C from high resolution near-infrared spectroscopic observations. *Astron. Astrophys.* **412**, 305–316. (2003)

Salo, H.: Simulations of dense planetary rings. III. Self-gravitating identical particles. Icarus 117, 287-312 (1995).

Sato, B., et al. : The N2K Consortium. II. A Transiting Hot Saturn around HD 149026 with a Large Dense Core. Astrophysical Journal 633, 465-473 (2005).

Saumon, D., Guillot, T. : Shock Compression of Deuterium and the Interiors of Jupiter and Saturn. Astrophysical Journal 609, 1170-1180 (2004)

Shepelyansky D.L., Pikovsky A.S., Schmidt J., Spahn F. Synchronization mechanism of sharp edges in rings of Saturn. Submitted to MNRAS.

Stevenson, D.J. : Interiors of the giant planets. Annual Review of Earth and Planetary Sciences 10, 257-295 (1982)

Stevenson, D.J., Salpeter, E.E. : The phase diagram and transport properties for hydrogen-helium fluid planets. Astrophysical Journal Supplement Series 35, 221-237 (1977)

Stewart, G. R., S. J. Robbins, and J. E. Colwell: Evidence for a primordial origin of Saturn's rings. 39th DPS meeting, abstract no. 7.06 (2007).

Thommes, E.W., Matsumura, S., Rasio, F.A.: Gas Disks to Gas Giants: Simulating the Birth of Planetary Systems. Science 321, 814 (2008)

Tiscareno M.S., Burns J.A., Hedman M.M., Porco C.C.: The population of propellers in Saturn's A ring. Astron. J. **135**, 1083-1091 (2008)

Toomre, A.: On the gravitational stability of a disk of stars. ApJ **139**, 1217-1238 (1964)

Tsiganis, K., Gomes, R., Morbidelli, A., Levison, H. F.: Origin of the orbital architecture of the giant planets of the Solar System. Nature 435, 459-461 (2005)

Weidenschilling, S.J., Lewis, J.S. (1973) Atmospheric and cloud structure of the Jovian planets. Icarus 20, 465–476.

Wuchterl G., Guillot T & Lissauer J.J.: The formation of giant planets, In Protostars & Planets IV, Tucson: University of Arizona Press; eds Mannings V., Boss, A.P., Russell, S. S., p. 1081 (2000)